\documentstyle[12pt,epsfig,amssymb]{article}

{\end{list}}
\newcounter{enumct}

\begin{document}
 
\sloppy

\thispagestyle{empty}
\begin{flushright}
\large
DTP/98/80 \\ November 1998
\end{flushright}
\vspace{0.85cm}

\begin{center}
\LARGE
{\bf The Parton Content of Virtual Photons}

\vspace{1.5cm}
\Large
M.\ Stratmann\\
\vspace{0.5cm}
\large
Department of Physics, University of Durham,\\ 
\vspace{0.1cm}
Durham DH1 3LE, England\\
\vspace{3.0cm}
{\bf Abstract} 
\end{center}
\vspace*{0.5cm}

\noindent
The QCD treatment of the parton structure of virtual photons is
briefly recalled, and possible limitations and open questions are
pointed out. Various models for these densities 
are compared, completed by a short discussion of
the treatment of heavy flavors. Finally, different ways to measure
the parton distributions of virtual photons in $e^+e^-$ and $ep$
experiments are summed up.
\normalsize

\vspace{7.4cm}
\noindent
{\it Talk presented at the workshop on
`Photon Interactions and the Photon Structure', Lund, Sweden,
September 1998.}
\vfill

\setcounter{page}{0}
\newpage

\begin{center}
{\LARGE\bf The Parton Content of Virtual Photons}\\
\vspace*{13mm}
{\Large Marco Stratmann} \\[3mm]
{\it Department of Physics, University of Durham, 
Durham DH1 3LE, England}\\[1mm]
{\it E-mail: Marco.Stratmann@durham.ac.uk}\\[16mm]

{\bf Abstract}\\[1mm]
\begin{minipage}[t]{140mm}
The QCD treatment of the parton structure of virtual photons is
briefly recalled, and possible limitations and open questions are
pointed out. Various models for these densities 
are compared, completed by a short discussion of
the treatment of heavy flavors. Finally, different ways to measure
the parton distributions of virtual photons in $e^+e^-$ and $ep$
experiments are summed up.
\end{minipage}\\[5mm]

\rule{160mm}{0.4mm}

\end{center}

%
\section{Introduction}
%
Theoretical studies of photonic parton distributions of {\em{real}}, i.e.,
on-shell, photons have a long history initiated by Witten's work 
\cite{ref:witten}.
On the experimental side the past few years have seen much 
progress since the advent of HERA.
The observation of `resolved' photon induced $ep$ processes, like
(di-)jet photoproduction, allows for tests of the hadronic nature of
(real) photons which are complementary to structure function
measurements in $e^+e^-$ collisions, where new results from LEP/LEP2
have improved our knowledge as well \cite{ref:newexp}.

Studies of the transition of the (di-)jet cross section from the
photoproduction to the deep-inelastic scattering (DIS) region at HERA 
point to the existence of a parton content also for {\em virtual} photons 
\cite{ref:expvirt,ref:maxfield}.
These measurements have revived the theoretical interest in this subject 
and have triggered a series of analyses of the dependence of
the $ep$ jet production cross section on the virtuality of the 
exchanged photon \cite{ref:grs2,ref:jets}. 
Recently, a next-to-leading order (NLO) QCD calculation of the
(di-)jet rate in $ep$ (and $e\gamma$) scattering, which
properly includes the contributions due to resolved virtual photons, 
has become available \cite{ref:kp,ref:jetvip}, and
resolved virtual photons have been included 
for the first time also in the Monte Carlo event generator {\tt RAPGAP}
\cite{ref:rapgap}.

Pioneering work on the parton structure of virtual photons has been
already performed a long time ago [10-12].
However, phenomenological models for these distributions 
have been proposed only in recent years [13-16] in view of the expected
experimental progress.
Ongoing measurements at HERA and future structure function measurements at 
LEP2 should seriously challenge these models and 
hopefully lead to a better understanding of the transition between 
the photoproduction and the DIS regime.
To finish this introductory prelude, let us stress that photons provide
us with a unique opportunity to investigate its parton content
in a {\em continuous} range of masses (virtualities) in contrast to
the situation with nucleons or pions.

The framework for parton distributions of virtual photons,
theoretical expectations, and open questions
are briefly recalled in Sec.~2. The various different models for the
parton content of virtual photons are compared in Sec.~3, 
supplemented by a short discussion of the treatment of heavy flavors
in Sec.~4. In Sec.~5 we sum up the different ways to measure the
parton densities of virtual photons in $ep$ and $e^+e^-$ 
experiments.
%
\section{Theoretical framework: definitions, expectations, open questions}
%
For clarity we henceforth denote the probed target photon with virtuality
$P^2=-p_{\gamma}^2$ by $\gamma (P^2)$, where $p_{\gamma}$ is the four
momentum of the photon emitted from, say, an electron in an $e^+ e^-$ or
$ep$ collider\footnote{In the latter case it is common to use $Q^2=-q^2$
instead of $P^2$, but we prefer $P^2$ according to the 
original notation used in $e^+e^-$ annihilations 
\cite{ref:uematsu,ref:rossi}, where it refers to 
the virtuality of the probed
(virtual) target photon, and $Q^2$ is reserved for the highly virtual
probe photon $\gamma^*(Q^2)$, $Q^2=-q^2\gg P^2$.}. For real 
$(P^2=0)$ photons we further simplify the notations by setting, as usual,
$\gamma \equiv \gamma (P^2=0)$.

The concept of photon structure functions for real and virtual 
({\em transverse})
photons can be defined and understood, in close analogy
to deep-inelastic lepton-nucleon scattering, via the subprocess
$\gamma^*(Q^2) \gamma(P^2) \rightarrow X$, as in $e^+e^-\rightarrow e^{\pm}X$
(`single tag') or $e^+e^-\rightarrow e^+e^- X$ (`double tag'). 
The relevant `single tag'
differential cross section can be expressed as in the hadronic case
in terms of the common scaling variables $x$ and $y$ 
\begin{equation}
\label{eq:eq1}
\frac{d^2\sigma(e\gamma(P^2)\rightarrow eX)}{dx dy}= 
\frac{2\pi\alpha^2 S_{e\gamma}}
{Q^4} \left[ \left(1+(1-y)^2\right)F_2^{\gamma (P^2)}(x,Q^2) - y^2
F_L^{\gamma (P^2)}(x,Q^2)\right]
\end{equation}
with $F_{2,L}^{\gamma (P^2)}$ denoting the photonic structure functions.
The measured $e^+e^-$ cross section is obtained by
convoluting (\ref{eq:eq1}) with the photon   
flux for the target photon $\gamma (P^2)$ \cite{ref:flux}.
The range of photon `masses' (virtualities) produced is
\begin{equation}
\label{eq:eq2}
m_e^2\; y^2/(1-y) \le P_{min}^2 \le P^2 \le P_{max}^2 \le \frac{S}{2}\;
(1-y) (1- \cos \Theta_{max})\;\;,
\end{equation}
where $y$ is the energy fraction taken by the photon $(y=E_{\gamma}/E_e)$, 
$S$ is the available squared c.m.s.\ energy, and $\Theta_{max}$ is the
maximal scattering angle of the electron in this
frame. $P^2_{min,max}$ in (\ref{eq:eq2}) are further determined by 
detector specifications and/or an eventual tagging of the outgoing
electron at the photon producing vertex.
$P^2_{min}$ effectively measures to a good 
approximation the dominant photon virtuality involved, just as 
$P^2_{min}=m_e^2y/(1-y)\simeq 0$ represents quasi-real photons. 
Even in the latter case there is, however, still a
small contribution from the high-$P^2$, virtual photon tail of the 
spectrum, which has to be estimated \cite{ref:dg,ref:aurenche}
when one tries to extract the parton densities of real photons.
For {\em transverse} virtual target photons $\gamma (P^2)$, whose virtuality
$P^2$ is essentially given by $P^2 \simeq P^2_{min}$, one expects 
\cite{ref:uematsu,ref:rossi} a 
parton content $f^{\gamma (P^2)} (x,Q^2)$ along similar lines as for
real photons. The range of applicability of this `picture', however, deserves
a further scrutiny.

For real photons $\gamma$ it is well-known that
in the framework of the quark parton
model (QPM) the $x$- and $Q^2$-dependence of 
$F_{2,L}^{\gamma}\equiv F_{2,L}^{\gamma (P^2)}$
is fully calculable from the `pointlike' QED process 
$\gamma^*(Q^2)\gamma \rightarrow q\bar{q}$ if one introduces
quark masses $m_q$ to regulate the mass singularities due to
$P^2=0$ \cite{ref:zerwas}.
However, this description is subject to perturbative QCD corrections
due to gluon radiation not present in the QPM \cite{ref:witten,ref:buras}.
The logarithmically enhanced contributions $\alpha_s \ln Q^2/Q_0^2$ 
can be resummed to all orders, removing the dependence on effective
quark masses, where $Q_0$ denotes some a priori {\em{not}} fixed 
renormalization point somewhere in the perturbative 
region $Q_0\gg\Lambda_{QCD}$.
Of course, this is not the whole story, since the 
photon can undergo a transition
into a vector meson of the same quantum numbers, which is afterwards probed
by the $\gamma^*(Q^2)$ (Vector
Meson Dominance (VMD) assumption). This {\em{non}}-perturbative part obeys
the same evolution equations as known from the hadronic case. 

Turning to {\em{virtual}} photons, i.e., $P^2\neq 0$, it is {\em{expected}}
\cite{ref:uematsu,ref:rossi} that for large enough virtualities $P^2$
one ends up with a {\em{fully perturbative}} prediction 
irrespective of $Q^2$. 
To facilitate the discussions, it is useful to define the relevant different 
ranges of $P^2$:
\begin{displaymath}
\renewcommand{\arraystretch}{1.5}
\begin{array}{cccccc}
(\mbox{I}) & P^2\ll\Lambda_{QCD}^2\ll Q^2&,& 
(\mbox{II}) & P^2\simeq \Lambda_{QCD}^2 &,\\
(\mbox{III}) & \Lambda_{QCD}^2 \ll P^2 \ll Q^2 &,&
(\mbox{IV}) & P^2 \simeq Q^2 &.
\end{array}
\end{displaymath}
Case (I) we have already discussed above, since it refers to a (quasi-)real
photon with $P^2\simeq 0$. In range (III) one can apply similar 
considerations as long as one restricts oneself to {\em{transverse}} 
virtual photons \cite{ref:uematsu,ref:rossi}, with the important 
distinction that $P^2$ is now within the perturbative domain and hence can
serve to fix $Q_0$, i.e., $Q_0={\cal{O}(P)}$.
This is the basis for the above mentioned conjecture of 
absolute predictability in this case, since any
non-perturbative VMD-inspired contributions are expected to vanish like
$(1/P^2)^2$ due to such a `dipole' suppression factor in the vector meson
`propagator'. 

Several questions have to be addressed: up to which values of 
$P^2$ (and $x$, $Q^2$) is the non-perturbative part relevant?
What are the lower and upper bounds on $P^2$ in (III), 
i.e., where and how takes the
transition to regions (II) and (IV), respectively, place, and 
down to which value of $P^2$ in (III) should one trust 
perturbation theory?
For smaller $P^2$, i.e., for a transition to the parton content
of real photons (I), one has to find some appropriate, physically
motivated prescription which {\em{smoothly}} extrapolates 
through region (II), where perturbation theory cannot be applied,
down to $P^2=0$.
On the other side, $P^2$ is bounded from above by $P^2\ll Q^2$ in order 
to avoid power-like (possibly higher twist) terms $(P^2/Q^2)^n$ 
which should spoil the dominance of the resummed logarithmic contributions 
$\sim \alpha_s \ln Q^2/P^2$ and, furthermore, to guarantee the dominance of 
the transverse photon contributions in physical cross sections.
For $P^2$ approaching $Q^2$ (region (IV)) the $e^+e^-$ result should 
reduce to the one given by the {\em{full}} fixed order
box $\gamma^*(Q^2)\gamma(P^2)\rightarrow q\bar{q}$ including all
$\left(P^2/Q^2\right)^n$ terms and possibly 
${\cal{O}}(\alpha_s)$ QCD corrections, which are unfortunately  
unknown so far. 

The question of when fixed order perturbation theory becomes 
the more reliable prescription
and the concept of virtual transverse photonic parton distributions 
(i.e., resummations) becomes irrelevant and perhaps misleading is 
in some sense similar 
to the question of whether heavy quarks should be 
treated as massless partons or not, which was extensively discussed
in the literature recently \cite{ref:grshq,ref:acot}.
Both issues are characterized by the appearance of at least two 
different, large scales, $P^2$ and $Q^2$ (or $m_q^2$ and $Q^2$), 
which might be indicative for resummations or not. 
In our case here, however, one is also interested in the 
transition to a region where resummations are indispensable 
(i.e., for real photons), 
but the range of applicability of this approach
with respect to $P^2$ (and possibly $x$ and $Q^2$) cannot be 
determined reliably so far unless the full NLO corrections to the 
$\gamma^*(Q^2)\gamma(P^2)$ box will be available 
to analyze its perturbative stability.

As already mentioned,
for a given $Q^2\gg P^2$ and increasing $P^2$ one expects 
that the resummed results approach the QPM result
determined for $m_q^2\ll P^2\ll Q^2$, due to the shrinkage of the evolution
length, i.e., less gluon radiation. 
The QPM result can be obtained from the process
$\gamma^*(Q^2)\gamma(P^2)\rightarrow q\bar{q}$, but now
$P^2\neq 0$ can act as the regulator and no quark masses have to be 
introduced. Taking the
limit $P^2/Q^2\rightarrow 0$ whenever possible, one obtains for
$F_2^{\gamma(P^2)}$ \cite{ref:uematsu,ref:rossi}
\begin{equation}
\label{eq:eq3}
\frac{1}{x} F_{2,QPM}^{\gamma (P^2)}(x,Q^2) =  3\sum_q e_q^4
\frac{\alpha}{\pi} \Bigg\{\left[x^2+(1-x)^2\right] 
\left( \ln \frac{Q^2}{P^2} + \ln \frac{1}{x^2}
\right) -2 +6x -6x^2\Bigg\} \;\;.
\end{equation}
It is important to notice that (\ref{eq:eq3}) 
is {\em different} from the result for on-shell $(P^2=0)$
photons \cite{ref:zerwas}, due to the different 
regularization adopted here\footnote{Note 
that $F_L^{\gamma (P^2)}$ is independent of the regularization 
adopted for calculating $\gamma^* (Q^2) \gamma (P^2)\rightarrow q\bar{q}$.}.
This difference will be relevant also for the
formulation of a model for the parton content of virtual photons, since
it is part of the perturbatively calculable boundary 
condition in NLO \cite{ref:uematsu,ref:rossi}.

The $Q^2$-evolutions of the photonic parton distributions
are essentially the same for real and virtual transverse photons. 
The inhomogeneous evolution equations are
most conveniently treated in the Mellin $n$ moment space,
where all convolutions simply factorize, and
the solutions can be given analytically (see, e.g.,
\cite{ref:disgamma,ref:grs}).
Let us only recall here that the distributions $f^{\gamma(P^2)}(x,Q^2)$,
obtained from solving the inhomogeneous evolution equations, can be 
separated into a `pointlike' (inhomogeneous) and a `hadronic'
(homogeneous) part
\begin{equation}
\label{eq:eq4}
f^{\gamma (P^2),n}(Q^2)= f_{PL}^{\gamma (P^2),n}(Q^2)+
f_{HAD}^{\gamma (P^2),n} (Q^2)\;\;.
\end{equation}
In NLO the pointlike singlet solution is schematically given by
\cite{ref:disgamma,ref:grs}
\begin{equation}
\label{eq:eq5}
\vec{f}_{PL}^{\,\gamma (P^2),n} = \left( \frac{2\pi}{\alpha_s}+\hat{U}\right)
\left(1-L^{1+\hat{d}\,}\right) \frac{1}{1+\hat{d}}\, \vec{a} +
\left(1-L^{\hat{d}\,}\right) \frac{1}{\hat{d}}\, \vec{b}\;\;,
\end{equation}
and the usual NLO hadronic solution can be found, e.g, in 
\cite{ref:disgamma,ref:grs}.
$\vec{a}$, $\vec{b}$, $\hat{d}$, and $\hat{U}$ in (\ref{eq:eq5}) stand
for certain combinations of the photon-parton splitting functions and
the QCD $\beta$-function \cite{ref:disgamma,ref:grs}, and 
$L\equiv \alpha_S(Q^2)/\alpha_s(Q_0^2)$.

Let us finish this technical part by quoting the relevant NLO expression
for the structure function $F_{2}^{\gamma (P^2)}(x,Q^2)$.
It should be pointed out that the treatment and 
expressions for $f^{\gamma (P^2)}(x,Q^2)$
(as {\em{on-shell}} transverse partons obeying the usual 
$Q^2$-evolution equations) presented above
{\em{dictates}} an identification of the relevant resolved
$f^{\gamma (P^2)} X \rightarrow X'$ sub-cross sections
with that of the real photon according to 
$\hat{\sigma}^{f^{\gamma (P^2)} X \rightarrow X'} =
\hat{\sigma}^{f^{\gamma} X \rightarrow X'}$.
In particular, the calculation of $F_2^{\gamma (P^2)}(x,Q^2)$ requires
the same hadronic Wilson coefficients $C_{2,q}$ and $C_{2,g}$ as for $P^2=0$,
\begin{equation}
\label{eq:eq6}
F_2^{\gamma (P^2)}\! =\!\!  \sum_{q=u,d,s} \!\! 2x e_q^2 \Bigg\{
q^{\gamma (P^2)}\! +\frac{\alpha_s}{2 \pi} 
\left(C_{2,q} \ast q^{\gamma (P^2)} +
      C_{2,g} \ast g^{\gamma (P^2)} \right)\! +
\frac{\alpha}{2\pi} e_q^2 C_{2,\gamma} \Bigg\}+ F_{2,c}^{\gamma (P^2)} \;,
\end{equation}
where $F_{2,c}^{\gamma (P^2)}$ represents the charm contribution (see
Sec.~4) and $\ast$ denotes the usual Mellin convolution.
Note that in the ${\rm{DIS}}_{\gamma}$ scheme, the NLO 
direct photon contribution $C_{2,\gamma}$ in (\ref{eq:eq6}) 
is absorbed into the evolution 
of the photonic quark densities, i.e., $C_{\gamma,2}=0$ \cite{ref:disgamma}.
The difference in the QPM expressions for $F_2^{\gamma (P^2)}$ between
real and virtual photons (as pointed out below Eq.~(\ref{eq:eq3})),
i.e., in the expressions for $C_{2,\gamma}$, is then accounted for by
a perturbatively calculable boundary condition 
for $q^{\gamma (P^2)}$ in NLO \cite{ref:grs}.
The LO expression for $F_2^{\gamma (P^2)}$ is obviously entailed in
(\ref{eq:eq6}) by dropping all $C_{i,\gamma}$.
Finally, it should be noted that $F_2^{\gamma (P^2)}$
is  kinematically constrained within \cite{ref:rossi} 
$0\le x\le (1+P^2/Q^2)^{-1}$.
%
\section{Comparison of different theoretical models}
%
Let us now briefly highlight the main features of the available
theoretical models for the parton densities of virtual photons:
\newline
\noindent
{\bf{GRS (Gl\"{u}ck, Reya, Stratmann) \cite{ref:grs}:}}
The GRS distributions provide a straightforward and simple extension
of the phenomenologically successful GRV photon densities \cite{ref:grvphoton}
to non-zero $P^2$ in LO {\em and} NLO. As for the GRV densities, the
NLO boundary conditions are formulated in the $\mathrm{DIS}_{\gamma}$
factorization scheme, originally introduced for $P^2=0$ 
to overcome perturbative instability problems arising in the conventional 
$\overline{\mathrm{MS}}$ scheme for large values of 
$x$ (see \cite{ref:disgamma} for details). 
At the low input scale $Q_0=\mu \simeq 0.5\,\mathrm{GeV}$,
universal for all `radiatively generated' GRV distributions
(proton, pion, and photon), the parton densities of
real photons are solely given by a simple VMD-inspired input
in LO and NLO($\mathrm{DIS}_{\gamma}$).
All one needs to fully specify the distributions for $P^2\ne 0$ is a 
simple, physically reasonable prescription which smoothly
interpolates between $P^2=0$ (region (I)) and $P^2\gg \Lambda_{QCD}^2$
(region (III)). This may be fixed by \cite{ref:grs} 
\begin{equation}
\label{eq:eq7}
f^{\gamma (P^2)}(x,Q^2=\tilde{P}^2) = \eta(P^2)
f_{non-pert}^{\gamma (P^2)}
(x,\tilde{P}^2) + \left[ 1- \eta(P^2)\right]
f_{pert}^{\gamma (P^2)}(x,\tilde{P}^2)
\end{equation}
with $\tilde{P}^2=\max(P^2,\mu^2)$ and
$\eta(P^2) = (1 +P^2/m_{\rho}^2)^{-2}$
where $m_{\rho}$ refers to some effective mass in the 
vector-meson propagator.
Note that the ansatz (\ref{eq:eq7}) implies that the input parton
distributions are frozen at the input scale $\mu$ for
real photons for $0\le P^2\le \mu^2$ such that the only 
$P^2$ dependence in region (II) stems from the dipole dampening
factor $\eta(P^2)$.
In NLO($\mathrm{DIS}_{\gamma}$) the perturbatively calculable 
input $f^{\gamma (P^2)}_{pert}(x,\tilde{P}^2)$ 
in Eq.~(\ref{eq:eq7}) is determined by the QPM box result (\ref{eq:eq3}); 
in LO it vanishes (see \cite{ref:grs} for details).
Since almost nothing is known experimentally about the parton structure
of vector mesons, the VMD-like non-perturbative input
is simply taken to be proportional to the 
GRV pion densities $f^{\pi}$ \cite{ref:grvpion}
\begin{equation}
\label{eq:eq8}
f_{non-pert}^{\gamma (P^2)}(x,\tilde{P}^2) =
\kappa\; (4\pi \alpha/f_{\rho}^2)
\times \left\{ \begin{array}{ccc}
f^{\pi}(x,P^2)& , &P^2>\mu^2 \\
& & \\
f^{\pi}(x,\mu^2) &,& 0\le P^2 \le \mu^2
\end{array} \right.
\end{equation}
where $\mu$, $\kappa$, $f_{\rho}$ are specified in \cite{ref:grvphoton}.

\begin{figure}[th]
\begin{center}
\vspace*{-0.6cm}
\epsfig{file=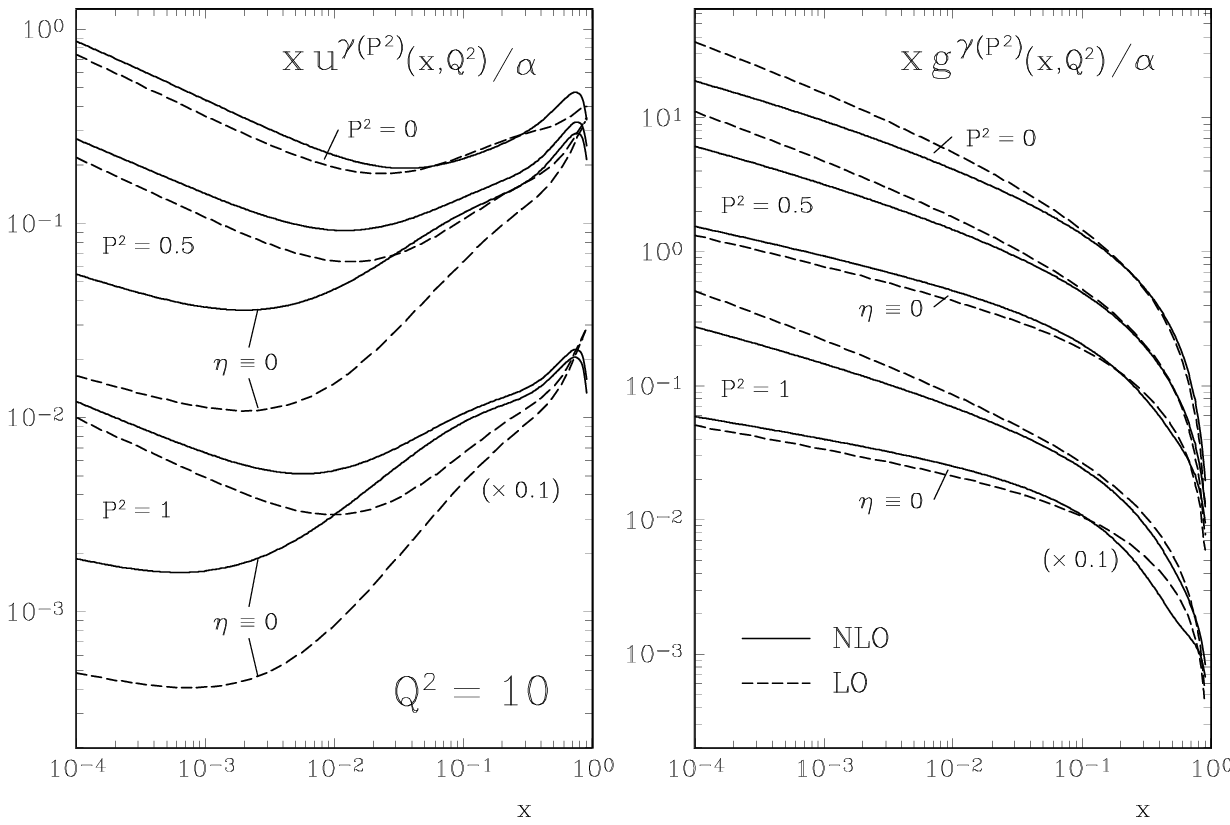,width=9cm,angle=90}
\caption{\label{viphfig7} \sf GRS \cite{ref:grs}
LO and NLO$(\mathrm{DIS}_{\gamma})$ 
predictions for the $u$-quark and gluon content
of a virtual photon for $Q^2=10\,\mathrm{GeV}^2$ 
and various fixed values of $P^2 ({\rm{GeV}}^2)$.
For comparison, the LO and NLO GRV parton distributions of the 
real photon $(P^2=0)$ \cite{ref:grvphoton} are shown as well.}
\end{center}
\vspace*{-0.7cm}
\begin{center}
\epsfig{file=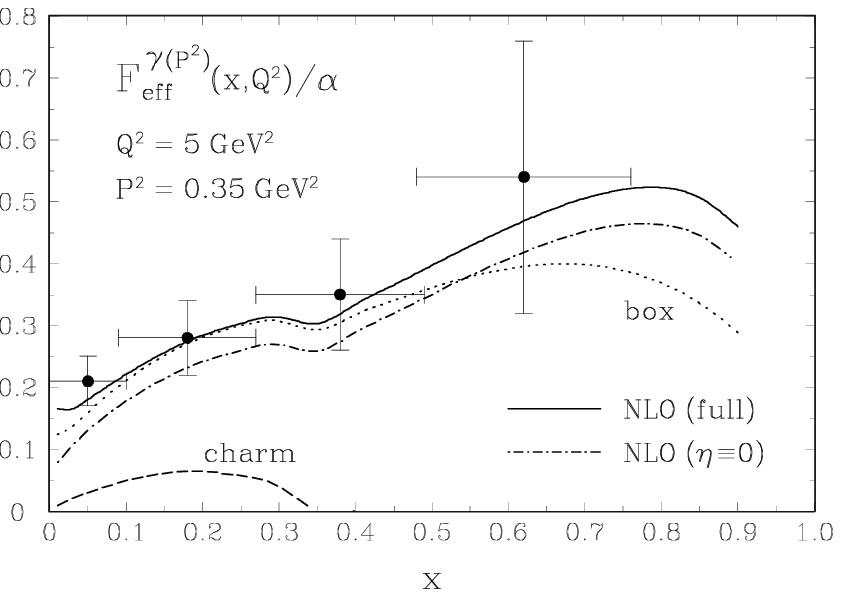,width=8.5cm,angle=270}
\caption{\label{viphfig5} \sf NLO GRS predictions for $F_{eff}^{\gamma (P^2)}
\equiv F_2+\frac{3}{2} F_L$ \cite{ref:grs}. 
The data points are taken from PLUTO \cite{ref:pluto}. The purely
perturbative results correspond to $\eta\equiv 0$ in Eq.~(\ref{eq:eq7}).}
\end{center}
\vspace*{-0.7cm}
\end{figure}
The resulting $u$-quark and gluon distributions
$u^{\gamma(P^2)}(x,Q^2)$ and $g^{\gamma(P^2)}(x,Q^2)$, respectively,
are shown in Fig.~1 for  $Q^2=10\,\mathrm{GeV}^2$ and
some representative values of $P^2$. 
With $g^{\gamma (P^2)}$ being the same
in the ${\rm{DIS}}_{\gamma}$ and $\overline{{\rm{MS}}}$
scheme, the shown ${\rm{DIS}}_{\gamma}$ results for $u^{\gamma (P^2)}$
can be easily transformed to the conventional $\overline{{\rm{MS}}}$ 
scheme \cite{ref:disgamma,ref:grvphoton,ref:grs}.
As can be inferred from the purely perturbative ($\eta \equiv 0$)
contributions, the non-perturbative components, entering for
$\eta \ne 0$ in Eq.~(\ref{eq:eq7}), are non-negligible and partly even
dominant (especially for $x\lesssim 0.01$).
It turns out \cite{ref:grs} that only for 
unexpectedly large $P^2\gg m_{\rho}^2$, say, $P^2 (\ll Q^2)$ larger 
than about $10\,{\rm{GeV}}^2$, 
the perturbative component starts to dominate over the 
entire $x$ range shown (cf., e.g., Fig.~13 in \cite{ref:grs}
for $P^2=20,\,100\,{\rm{GeV}}^2$ and $Q^2=1000\,{\rm{GeV}}^2$).

The precise form of $\eta(P^2)$ in Eq.~(\ref{eq:eq7})
clearly represents, apart from $f_{non-pert}^{\gamma (P^2)}$ itself, the
largest uncertainty in this model and has to be tested by future experiments.
The only measurement of the virtual photon structure in 
$\gamma^*(Q^2) \gamma(P^2)$ DIS available thus far \cite{ref:pluto}, 
is compared in Fig.~2 with the NLO GRS 
prediction for $F_{eff}^{\gamma (P^2)}\equiv F_2+\frac{3}{2} F_L$,
the combination measured effectively by PLUTO \cite{ref:pluto}
(see also Sec.~5). Due to the poor statistics of the data and the 
rather limited $x$ range, the resummed NLO result cannot be
distinguished from the naive, not resummed QPM result (dotted curve).

Finally, it should be noted that in the GRS approach heavy quarks
do not take part in the $Q^2$-evolution, i.e., there is {\em no}
`massless' photonic charm distribution \cite{ref:grs}. 
Heavy flavors can only be produced {\em extrinsically}, 
and their contributions have to be calculated 
according to the appropriate massive sub-cross sections (see also Sec.~4).
The LO GRS distributions are available in parametrized form for
$P^2<10\,\mathrm{GeV}^2$ and $Q^2\gtrsim 5 P^2$ \cite{ref:grs2}.
\newline
\noindent
{\bf{SaS (Schuler and Sj\"{o}strand) \cite{ref:sas1,ref:sas2}:}} 
The starting point for their analysis are also some well-established
LO sets of parton densities for real photons \cite{ref:sas1}, 
SaS 1D and SaS 2D, corresponding to two rather different  
assumptions about the non-perturbative hadronic input\footnote{The
additional SaS 1M and SaS 2M sets \cite{ref:sas1} are theoretically
inconsistent, as the LO evolved densities are combined 
with the NLO scheme-dependent photon-coefficient function $C_{2,\gamma}$
in the calculation of $F_2^{\gamma}$ in LO. These sets should not be used in
phenomenological analyses.}.   
The SaS 1D set has a similarly low input scale $Q_0\simeq0.6\,\mathrm{GeV}$
as in the GRV \cite{ref:grvphoton} and GRS \cite{ref:grs} analyses, but
instead of simply relating the VMD input distributions to that of a
pion, a fit is performed to the coherent sum of the lowest-lying
vector meson states $\rho, \omega, \phi$.
For the SaS 2D set a `conventional' high input scale $Q_0=2\,\mathrm{GeV}$
is used at the expense of two additional fit parameters, one characterizing
the necessary additional `hard' component for the quark input at larger values
of $Q_0$, the other models the effect of additional vector meson states
beside the ones already taken into account in the SaS 1D set.
The shape of the SaS gluon densities are entirely fixed by 
theoretical estimates, no direct or indirect constraints 
from direct-photon production data in $\pi p$ collisions as in
the GRV analysis \cite{ref:grvphoton} have been imposed.
Contrary to GRS, heavy flavors are included as massless partons in
the photon above the threshold $Q^2>m^2_q$, but when calculating/fitting 
the available $F_2^{\gamma}(x,Q^2)$ data, the massive 
`Bethe-Heitler' cross section for $\gamma^*\gamma\rightarrow c\bar{c}$ 
is used instead, which is, however, 
not entirely consistent due to double counting.

The extension to non-zero $P^2$ is based on the fact that the $n$ moments
of the photon densities parton can be expressed as a
dispersion-integral in the mass $k^2$ of the $\gamma \rightarrow q\bar{q}$
fluctuations, which links perturbative and non-perturbative contributions
\cite{ref:sas1,ref:sas2}.
Having assumed some model-dependent weight function for the 
dispersion-integral, and after associating the low-$k^2$ part 
with some discrete set of vector mesons (as for $P^2=0$),
one arrives at their final expression
for the parton densities of virtual photons \cite{ref:sas1,ref:sas2}
\begin{equation}
\label{eq:sas1}
f^{\gamma (P^2)}(x,Q^2) \!=\!\! \sum_V \frac{4\pi \alpha}{f_V^2}
\left[ \frac{m_V^2}{m_V^2+P^2}\right]^2 \! f^{\gamma,V}(x,Q^2,\tilde{P^2}) +
\sum_q \frac{\alpha}{\pi} e_q^2 \int_{\tilde{P}^2}^{Q^2} \!\!
\frac{dk^2}{k^2} f^{\gamma\rightarrow q\bar{q}}(x,Q^2,k^2)\;,
\end{equation}
where in the perturbative contribution the suppression factor
$[k^2/(k^2+P^2)]^2$ has been substituted by an effective lower
cut-off $\tilde{P}^2$ for the integration.
Both components $f^{\gamma,V}$ and $f^{\gamma\rightarrow q\bar{q}}$
in (\ref{eq:sas1}) integrate to unit momentum.
\begin{figure}[th]
\begin{center}
\vspace*{-1.5cm}
\epsfig{file=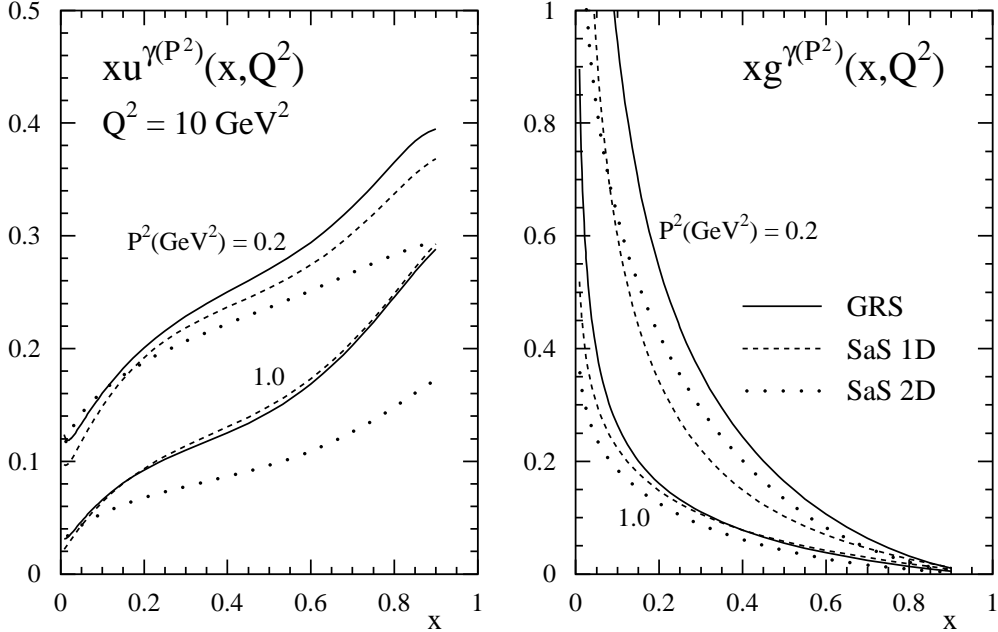,width=15cm}
\vspace*{-1cm}
\caption{\sf Comparison of the LO GRS predictions \cite{ref:grs} for the
$u$-quark and gluon content of virtual photons with the 
SaS 1D and SaS 2D results for $\tilde{P}^2=\max (Q_0^2, P^2)$
\cite{ref:sas1} at $Q^2=10\,{\rm{GeV}}^2$ and for two values of $P^2$.}
\end{center}
\vspace*{-0.7cm}
\end{figure}
As in the GRS model above, the VMD part contains a dipole suppression
factor, which dampens all non-perturbative contributions with
increasing virtuality $P^2$. Contrary the the GRS approach, several
different choices for the input scale $\tilde{P}^2$ have been studied 
apart from $\tilde{P}^2=\max (Q_0^2, P^2)$ \cite{ref:sas2}.
The differences between all these procedures can be viewed as a
measure for the theoretical uncertainty within this approach
(see, e.g., Figs.~1 and 2 in \cite{ref:sas2}).
It should be also noted that for some more complicated choices for 
$\tilde{P}^2$, the photonic parton densities obey evolution equations different
from those of the real photon, e.g., the inhomogeneous
term can be modified by a factor $Q^2/(Q^2+P^2)$ \cite{ref:sas2}.
All different sets of distributions are available in parametrized form
\cite{ref:sas1,ref:sas2} but, for the time being, 
the SaS analysis is restricted to LO only.

Fig.~3 compares the LO GRS with the LO SaS 1D and SaS 2D distributions
(choosing $\tilde{P}^2=\max (Q_0^2, P^2)$)
for $Q^2=10\,{\rm{GeV}}^2$ and two $P^2$ values relevant for future LEP2
measurements \cite{ref:gamgam}.
As can be seen, the SaS 1D results, which refer to a equally 
low input scale as used in GRS, and the GRS densities
are rather similar, at least for the $u$-quark, 
whereas the SaS 2D (quark) distributions are sizeably smaller
in this kinematical range, mainly due to the higher input scale.
For smaller values of $x$ (not shown in Fig.~3) the GRS 
densities rise more strongly than the SaS densities, due the different
non-perturbative input.  
However, with increasing $P^2$ all differences get, of course, more or less
washed out, since one approaches the purely perturbative domain and the 
differences in the treatment of the non-perturbative component 
become negligible.
\newline
\noindent
{\bf{DG (Drees and Godbole) \cite{ref:dg}:}}
The aim of this analysis is to estimate the influence of the
high-$P^2$ tail in the photon flux in untagged or anti-tagged 
events, i.e., to study the impact of virtual photons in a sample of almost 
real photons. This effect should be taken into account if one tries to
extract the parton densities of real photons from such measurements.
To perform a quantitative analysis, DG provide a simple interpolating 
multiplicative factor $r$ which can be applied to 
{\em any} set of distributions of real photons.
Several different forms for $r$ are studied in \cite{ref:dg}, 
and one of the main alternatives is
\begin{equation}
\label{eq:dg1}
r=1-\frac{\ln (1+P^2/P_c^2)}{\ln (1+Q^2/P_c^2)}\;\;,
\end{equation}
where $P_c$ denotes some typical hadronic scale like, for
instance, the $\rho$ mass.
The factor $r$ is applied to all quark flavors, but the gluon is 
expected to be further suppressed, since it is radiated off the quarks
\cite{ref:borz}:
\begin{equation}
\label{eq:dg2}
q^{\gamma (P^2)}(x,Q^2) = r q^{\gamma}(x,Q^2)\,\,,\,\,
g^{\gamma (P^2)}(x,Q^2) = r^2 g^{\gamma}(x,Q^2)\;\;.
\end{equation}
For typical $\gamma\gamma$ experiments with $Q^2\simeq 10\,\mathrm{GeV}^2$ 
in an untagged situation, virtual photon effects then suppress the
effective photonic quark and gluon content by 
about 10 and $15\%$, respectively \cite{ref:dg}.
 
Obviously, the above ansatz (\ref{eq:dg2}) does not change the $x$ shape of the
distributions, which would require more complicated forms for
$r$ \cite{ref:dg}. The recipe (\ref{eq:dg2}) does not appear 
to be well suited for QCD tests of the virtual photon content: 
on the one hand the densities
in (\ref{eq:dg2}) are not a solution of the inhomogeneous evolution
equations, and on the other hand there is no dipole power suppression factor
for the non-perturbative VMD part of the virtual photon densities as
in GRS \cite{ref:grs} or SaS \cite{ref:sas1,ref:sas2}, i.e., the
approximation (\ref{eq:dg2}) can only be applied in the large $x$ region
where the perturbative pointlike part dominates, whereas at smaller
$x$ it may grossly overestimates the densities for increasing virtuality
$P^2$.
A similar strategy as in Eqs.~(\ref{eq:dg1}) and (\ref{eq:dg2}) has
been used in \cite{ref:aurenche} including, however, a power 
suppressed VMD part. 
%
\section{Treatment of heavy flavors}
%
The question of how to treat heavy flavor $(m_q\gg\Lambda_{QCD})$
contributions to structure functions and cross sections in the most
appropriate and reliable way
has attracted a considerable amount of interest 
in the past few years \cite{ref:grshq,ref:acot,ref:dghq}.
This was mainly triggered by the observation that the charm contribution
to the DIS proton structure function $F_2^p$ amounts to about $20-25\%$
in the small $x$ region covered by HERA.
In case of the photon structure function $F_2^{\gamma (P^2)}$
in (\ref{eq:eq6}), effects due to 
charm are sizeable also in the large $x$ region due to the existence of 
the direct/pointlike component, such that a proper treatment is even
more important here.

There are two extreme ways to handle heavy flavors: one can simply
include heavy flavors as massless partons in the evolution above
some threshold $Q^2\gtrsim m_q^2$, or one can stick to a picture with
only light partons in the proton/photon. In the latter case, heavy
flavors do not participate in the evolution equations at all and can be
produced only extrinsically.
In case of $F_{2,c}^{\gamma (P^2)}$ in (\ref{eq:eq6}), 
two different contributions have
to be taken into account. Firstly, the direct `Bethe-Heitler' process
$\gamma^*(Q^2)\gamma(P^2)\rightarrow c \bar{c}$, and secondly the
resolved contribution $\gamma^*(Q^2)g^{\gamma(P^2)}\rightarrow c \bar{c}$.
For real photons these cross sections are known up to NLO and can be found 
in \cite{ref:hq1}. In \cite{ref:hq2} it was shown
that the two contributions are
separated in the variable $x$: for large $x$, say, $x\gtrsim 0.05$,
the direct process dominates, whereas for $x\lesssim 0.01$ the dominant
contribution stems from the resolved part.

The fully massless treatment has been abandoned recently in all 
existing modern sets of proton densities, simply because it does not exhibits 
the correct $x$ {\em and} $Q^2$ dependent threshold behavior.
A potential problem with the massive treatment are possibly large 
logarithms in the relevant sub-cross sections far above the threshold, 
which {\em might} be indicative for resummations, i.e., for introducing a
`massless' heavy quark distribution. 
Therefore various `unified' prescription were proposed recently 
\cite{ref:acot} which
reduce to the massive results close to, and to the massless 
picture far above threshold. For the time being these studies have been 
performed only for the proton densities and not in the context of photons.

However, as mentioned in Sec.~3, GRS \cite{ref:grs} 
prefer to stick to the fully massive framework, 
similarly to the case of the GRV proton densities \cite{ref:grvproton}.
In each case this is motivated by the observation that all relevant
fully massive production mechanisms appear to be perturbatively 
stable, at least for all experimentally relevant values of $x$ and $Q^2$
(see, e.g., Refs.~\cite{ref:grshq,ref:hq2}).
Moreover, all theoretical uncertainties, in particular the
dependence on the factorization scale appear to be  
well under control, leading to
the conclusion that there is no real need for any resummation procedure.
It should be mentioned that the relevant expressions for 
$\gamma^*(Q^2)\gamma(P^2)\rightarrow c \bar{c}$ for non-zero $P^2$ 
are available only in LO so far \cite{ref:budnev}, hence a study of the 
perturbative stability cannot be performed here yet.

Clearly, the fully massive treatment is much more cumbersome and
inconvenient than the massless or `unified' framework when 
calculating, for instance, jet production cross sections in $ep$
or $\gamma\gamma$ collisions. One cannot simply increase the number of
active flavors by one unit and use a $c^{\gamma (P^2)}(x,Q^2)$ distribution.
Instead one has to calculate the relevant sub-cross sections with massive
quarks for the final state configuration under consideration, which
is much more involved and time-consuming in numerical analyses. 
However, for large-$p_T$
jet production in LO ($m_q/p_T\ll 1$), for instance, one can simply
approximate the relevant massive cross sections by their massless
counterparts, neglecting of course all massless contributions with a
`heavy' quark in the initial state.
%
\section{Measuring $\gamma^*$-PDF's in $e^+e^-$ and $ep$ reactions}
%
Since there are several dedicated contributions which discuss 
recent experimental progress or future prospects
\cite{ref:maxfield,ref:kp,ref:jung}, we can be fairly
brief here and concentrate only on topics not covered elsewhere.
 
Let us first of all delineate the $x,\, Q^2,$ and $P^2$
ranges covered by LEP2 and the expected statistical accuracy for the virtual
photon structure functions measurements \cite{ref:gamgam}.
Because of its higher energy and
integrated luminosity, LEP2 can provide improved information from double
tagged events as compared to the not very precise 
results from PLUTO \cite{ref:pluto} shown in Fig.~2.
The most important double tag sample at LEP2 is expected to
come from events with $Q^2\gtrsim 3\,{\rm{GeV}}^2$ and
$0.1\lesssim P^2\lesssim 1\,{\rm{GeV}}^2$. For typically expected
$500\,\mbox{pb}^{-1}$ of data collected, about 800 of such events will be seen,
covering $3\cdot 10^{-4}\lesssim x<1$ and $3\lesssim Q^2 \lesssim
1000\,{\rm{GeV}}^2$ \cite{ref:gamgam}. The yield of events with both $Q^2$ and
$P^2 \gtrsim 3\,{\rm{GeV}}^2$ seems to be too small for a meaningful analysis.

Fig.~4 shows the virtual photon structure function
$F^{\gamma (P^2)}_2(x,Q^2)$ in LO as predicted by the 
SaS 1D, SaS 2D (using $\tilde{P}^2=\max (Q^2,P^2)$), and 
GRS models in two bins for $P^2$ and $Q^2$. 
The error bars indicate the statistical precision 
expected for each $x$ bin using the SaS 1D densities 
(similar for the SaS 2D and GRS distributions). A measurement of
$F_2^{\gamma (P^2)}(x,Q^2)$ within these bins, as distinct from the real
$(P^2=0)$ photon structure function (illustrated by the solid curves for
SaS 1D) should be possible at LEP2 and could be compared to 
the different model predictions, which turn out to be 
rather similar in the accessible $x,\,Q^2,$ and $P^2$ bins, except for 
the smallest $x$ bins and $P^2\rightarrow 0$ 
($\langle P^2\rangle =0.2\,{\rm{GeV}}^2$). 
The latter differences are of course related to the present ignorance 
of the $P^2=0$ distributions for $x\rightarrow 0$, i.e., 
whether they either steeply rise as in case of GRV \cite{ref:grvphoton}
or show a rather flat $x\rightarrow 0$ behaviour as, e.g., 
in case of SaS 2D \cite{ref:sas1,ref:sas2}.
%
\begin{figure}[th]
\vspace*{-1.5cm}
\begin{center}
\epsfig{file=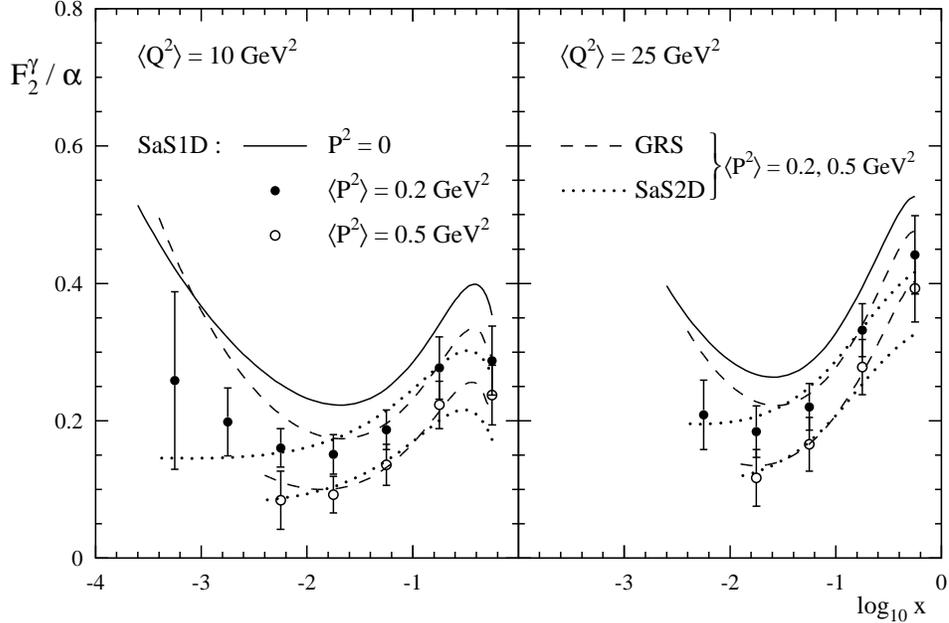,width=13.5cm}
\vspace*{-0.4cm}
\caption{\label{vphappfig1}\sf Expectations for the statistical accuracy
of the virtual photon structure measurement at LEP2 in two different
$P^2$ and $Q^2$ bins using the SaS 1D distributions \cite{ref:sas1}. 
The SaS 1D predictions for $P^2=0$ and the results for the GRS 
\cite{ref:grs} and SaS 2D models are shown as lines for comparison. 
The upper (lower) curves for GRS and SaS 2D refer to 
$P^2=0.2\, (0.5)\,{\rm{GeV}}^2$, respectively.
The figure is taken from \cite{ref:gamgam}.}
\end{center}
\vspace*{-0.7cm}
\end{figure}

However, apart from the experimental challenge there is an
additional complication already noticed by PLUTO \cite{ref:pluto}:
what is directly measured is, of course, 
{\em not} $F_2^{\gamma (P^2)}$ but the 
$\gamma^*(Q^2)\gamma(P^2)$ DIS cross section, which can be schematically
expanded as
$\sigma_{\gamma^*(Q^2)\gamma(P^2)}=\sigma_{TT}+\varepsilon_1 
\sigma_{LT} + \varepsilon_2 \sigma_{TL} + \varepsilon_1 \varepsilon_2
\sigma_{LL}$, where $L$ and $T$ denote longitudinal and transverse
polarization, respectively, of the probe and the target photons, and
$\varepsilon_{1,2}$ are the $L/T$ $\gamma$-flux ratios.
For PLUTO \cite{ref:pluto} $\varepsilon_1\simeq \varepsilon_2\simeq 1$
$(\Leftrightarrow y\ll1)$ and {\em assuming} that
$\sigma_{LL}\simeq 0$ and $\sigma_{LT}=\sigma_{TL}$, as for the
QPM expressions for $\gamma^*_{T,L}(Q^2)\gamma_{T,L}(P^2)\rightarrow q\bar{q}$
for vanishing constituent quark masses \cite{ref:budnev}, 
one arrives at the combination \cite{ref:pluto}
$\sigma_{\gamma^*\gamma}\sim F_2 +3/2 F_L
\equiv F_{eff}^{\gamma (P^2)}$ effectively measured by PLUTO (cf.\ Fig.~2).
Hence, strictly speaking such measurements cannot be directly related
to the densities $f^{\gamma(P^2)}(x,Q^2)$, since only 
{\em transverse} $(T)$ virtual photons are described by the GRS, SaS,
and DG models. Furthermore, in the QPM model \cite{ref:budnev} 
it turns out that the
contribution due to longitudinal target photons is rather sizeable
at large $x$ even for $P^2/Q^2\ll1$, contrary to the expectation that
transverse photons should dominate in this region. Clearly, more work is
required here for a meaningful interpretation of any future results 
from LEP2, possibly including also studies of the parton content of 
longitudinal photons which have not been carried out so far.

In $ep$ collisions (di-)jet production is certainly the best tool
to decipher the parton structure of virtual photons, and a lot of
experimental and theoretical progress was reported at the workshop
\cite{ref:maxfield,ref:kp}.
There should be a hierarchy between the hard scale $\mu_f^2$
($=Q^2$ in $e^+e^-$) at which the virtual photon is probed 
(typically $\mu_f^2={\cal{O}}(P^2+p_{T,jet}^2)$ in case of jet 
production) and the photon virtuality $P^2$ (see footnote 1).
Exactly in this kinematical domain an excess in the dijet rate was 
observed by H1 \cite{ref:maxfield}, which can be 
nicely attributed to a resolved virtual photon contribution, 
in accordance with all existing models for the $f^{\gamma (P^2)}$ 
described in Sec.~3.

Recently, similar studies were extended to 
the production rate of forward jets \cite{ref:jung},
which is regarded as a test in favor of the BFKL dynamics
\cite{ref:mueller}. Indeed the
usual DGLAP (direct photon induced) cross section falls short 
of the data by roughly a factor of two \cite{ref:h1forward}. 
In \cite{ref:jung} it was demonstrated,
however, that the inclusion of the resolved virtual photon component
removes this discrepancy, and the full DGLAP results are then
in even better agreement with data than the BFKL results, in particular for
the two to one forward jet ratio \cite{ref:h1forward}.
However one should be cautious to jump to any conclusions.
In order to suppress the phase space for the DGLAP evolution, the
$p_T^2$ of the forward jet is required to be of the same size as
the virtuality $P^2$ of the photon, hence there is no real
hierarchy between the hard scale $\mu_f^2={\cal{O}}(P^2+p_{T,jet}^2)$
and $P^2$ and thus no large logarithm $\ln \mu_f^2/P^2$ which
is resummed in $f^{\gamma (P^2)}$. Naively one would therefore expect only a
small resolved contribution, as was also
observed in the H1 jet analysis \cite{ref:maxfield} or in the 
theoretical studies \cite{ref:kp} for $P^2\rightarrow \mu_f^2$,
rather than a gain by about a factor of two.
Hence the kinematics of forward jets seems to be very subtle 
(for instance, the virtual photon content is only probed at large
momentum fractions $x_{\gamma}$), and
presumably the theoretical uncertainties due to scale variations and
changes in the model for the $f^{\gamma (P^2)}$ (in particular, of the input
scale $\tilde{P}^2$) are of the same size as the resolved photon
contribution itself. More detailed studies are clearly required here. 
Furthermore, it should be kept in mind that
all BFKL results so far are based only LO parton level calculations.
It seems, however, that the forward jet kinematics is not suited to 
distinguish between BFKL and DGLAP at HERA \cite{ref:jung}. 
%

%

\begin{thebibliography}{99}
%
\bibitem{ref:witten} E.\ Witten, Nucl. Phys. {\bf{B120}} (1977) 189.
%
\bibitem{ref:newexp} Recent experimental results can be found in, e.g., 
the proc.\ of the `Photon '97 Conference', 
Egmond aan Zee, 1997, A.\ Buijs and 
F.C.\ Erne (eds.); see also the contributions by
M.\ Kienzle, B.\ Surrow, I.\ Tyapkin, and Y.\ Yamazaki 
in these proceedings.
%
\bibitem{ref:expvirt} M.L. Utley, in
proc.\ of the `Europhysics Conference (HEP'95)', Brussels, 1995, 
J.\ Lemonne et al. (eds.), World Scientific, p.~570;\\
C.\ Adloff et al., H1 collab., Phys. Lett. {\bf B415} (1997) 418;
{\tt hep-ex/9806029}.
%
\bibitem{ref:maxfield} S.\ Maxfield, these proceedings.
%
\bibitem{ref:grs2} M.\ Gl\"{u}ck, E.\ Reya, and M.\ Stratmann,
Phys. Rev. {\bf D54} (1996) 5515.
%
\bibitem{ref:jets} D.\ de Florian, C.\ Garcia Canal, and R.\ Sassot,
Z. Phys. {\bf C75} (1997) 265;\\
J. Chyla and J.\ Cvach, in proc.\ of the 1995/96 workshop on
`Future Physics at HERA', DESY, G.\ Ingelman et al.\ 
(eds.), p.\ 545; in 
proc.\ of `Photon '97 Conference', Egmond aan Zee,
1997, A.\ Buijs and F.C.\ Erne (eds.).
%
\bibitem{ref:kp} M.\ Klasen, G.\ Kramer, and B.\ P\"{o}tter,
Eur. Phys. J. {\bf C1} (1998) 261;\\
G.\ Kramer and B.\ P\"{o}tter, Eur. Phys. J. {\bf C5} (1998) 665;\\
B.\ P\"{o}tter, these proceedings.
%
\bibitem{ref:jetvip} B.\ P\"{o}tter, {\tt hep-ph/9806437}. 
%
\bibitem{ref:rapgap} H.\ Jung, Comp. Phys. Comm. {\bf 86} (1995) 147.
%
\bibitem{ref:uematsu} T.\ Uematsu and T.F.\ Walsh, 
Phys. Lett. {\bf{101B}} (1981) 263, Nucl. Phys. {\bf{B199}} (1982) 93.
%
\bibitem{ref:rossi}  G.\ Rossi, Phys. Rev. {\bf{D29}} (1984) 852; UC San Diego
report UCSD-10P10-227 (unpublished).
%
\bibitem{ref:borz} F.M.\ Borzumati and G.A.\ Schuler, Z. Phys. {\bf{C58}}
(1993) 139.
%
\bibitem{ref:grs} M.\ Gl\"{u}ck, E.\ Reya, and M.\ Stratmann,
Phys. Rev. {\bf D51} (1995) 3220.
%
\bibitem{ref:sas1} G.A.\ Schuler and T.\ Sj\"{o}strand, 
Z. Phys. {\bf C68} (1995) 607.
%
\bibitem{ref:sas2} G.A.\ Schuler and T.\ Sj\"{o}strand,
Phys. Lett. {\bf B376} (1996) 193.
%
\bibitem{ref:dg} M.\ Drees and R.M.\ Godbole, Phys. Rev. {\bf D50} (1994)
3124.
%
\bibitem{ref:flux} C.F.\ von Weizs\"{a}cker, Z. Phys. {\bf 88} (1934) 612;\\
E.J.\ Williams, Phys. Rev. {\bf 45} (1934) 729;\\
S.\ Frixione, M.L.\ Mangano, P.\ Nason, and G.\ Ridolfi, 
Phys. Lett. {\bf B319} (1993) 339. 	
%
\bibitem{ref:aurenche} P.\ Aurenche et al., Prog. Theor. Phys. {\bf 92}
(1994) 175; in proc.\ of the workshop on `Two-Photon Physics at LEP
and HERA', Lund, 1994, G.\ Jarlskog and L.\ J\"{o}nsson (eds.),
p.\ 269.
%
\bibitem{ref:zerwas} T.F.\ Walsh and P.M.\ Zerwas, 
Phys. Lett. {\bf 44B} (1973) 195.
%
\bibitem{ref:buras} W.A.\ Bardeen and A.J.\ Buras, 
Phys. Rev. {\bf D20} (1979) 166; {\bf D21} (1980) 2041(E).
%
\bibitem{ref:grshq} M.\ Gl\"{u}ck, E.\ Reya, and M.\ Stratmann,
Nucl. Phys. {\bf B422} (1994) 37.
%
\bibitem{ref:acot} M.\ Aivazis, F.\ Olness, and W.-K.\ Tung,
Phys. Rev. {\bf D50} (1994) 3085;\\
M.\ Aivazis, J.C.\ Collins, F.\ Olness, and W.-K.\ Tung,
Phys. Rev. {\bf D50} (1994) 3102;\\
A.D.\ Martin, R.G.\ Roberts, M.G.\ Ryskin, and W.J.\ Stirling,
Eur. Phys. J. {\bf C2} (1998) 287;\\
R.S.\ Thorne and R.G.\ Roberts, Phys. Rev. {\bf D57} (1998) 6781;
Phys. Lett. {\bf B421} (1998) 303;\\
M.\ Buza, Y.\ Matiounine, J.\ Smith, and W.L.\ van Neerven,
Phys. Lett. {\bf B411} (1997) 211; 
Eur. Phys. J. {\bf C1} (1998) 301.
%
\bibitem{ref:disgamma} M.\ Gl\"{u}ck, E.\ Reya, and A.\ Vogt,
Phys. Rev. {\bf D45} (1992) 3986.
%
\bibitem{ref:grvphoton} M.\ Gl\"{u}ck, E.\ Reya, and A.\ Vogt,
Phys. Rev. {\bf D46} (1992) 1973.
%
\bibitem{ref:grvpion}  M.\ Gl\"{u}ck, E.\ Reya, and A.\ Vogt,
Z. Phys. {\bf C53} (1992) 651.
%
\bibitem{ref:pluto} Ch.\ Berger et al., PLUTO collab., Phys. Lett. {\bf 142B}
(1984) 119.
%
\bibitem{ref:gamgam} `$\gamma\gamma$ Physics' working group report,
P.\ Aurenche and G.A.\ Schuler (conv.), CERN 96-01, vol.~1, p.~291.
%
\bibitem{ref:dghq} M.\ Drees and R.M.\ Godbole, in proc.\ of the
`Photon '95 Conference', Sheffield, 1995, D.J.\ Miller et al. (eds.),
World Scientific, p.~123.
%
\bibitem{ref:hq1} E.\ Laenen, S.\ Riemersma, J.\ Smith, and
W.L.\ van Neerven, Phys. Rev. {\bf D49} (1994) 5753.
%
\bibitem{ref:hq2} E.\ Laenen and S.\ Riemersma, 
Phys. Lett. {\bf B376} (1996) 169.
%
\bibitem{ref:grvproton} M.\ Gl\"{u}ck, E.\ Reya, and A.\ Vogt,
Z. Phys. {\bf C67} (1995) 433; Eur. Phys. J. {\bf C5} (1998) 461.
%
\bibitem{ref:budnev} V.M.\ Budnev, I.F.\ Ginzburg, G.V.\ Meledin,
and V.G.\ Serbo, Phys. Rep. {\bf 15} (1975) 181.
%
\bibitem{ref:jung} H.\ Jung, L.\ J\"{o}nsson, and H.\ K\"{u}ster,
{\tt hep-ph/9805396};\\
H.\ Jung, these proceedings.
%
\bibitem{ref:mueller} A.H.\ Mueller, Nucl. Phys. Proc. Suppl. 
{\bf 18C} (1990) 125; J. Phys. {\bf G17} (1991) 1443.
%
\bibitem{ref:h1forward} J.\ Breitweg et al., ZEUS collab.,
{\tt hep-ex/9805016} (Eur. Phys. J. {\bf C}); \\
C.\ Adloff et al., H1 collab., {\tt hep-ex/9809028}.
%
\end{thebibliography}
\end{document}